\documentclass[twocolumn,showpacs,preprintnumbers,amsmath,amssymb,prb]{revtex4}

\usepackage{graphicx}
\usepackage{dcolumn}
\usepackage{bm}
\usepackage{amsmath}

\usepackage[usenames]{color}

\newcommand{\dd}{\mathrm{d}}
\newcommand{\rem}[1]{}
\newcommand{\qav}[1]{\left\langle #1\right\rangle}
\newcommand{\refe}[1]{(\ref{#1})}

\newcommand{\refE}[1]{Eq.~(\ref{#1})}
\newcommand{\beq}{\begin{equation}}
\newcommand{\eeq}{\end{equation}}
\newcommand{\beqa}{\begin{eqnarray}}
\newcommand{\eeqa}{\end{eqnarray}}

\renewcommand{\Re}[1]{{\rm Re}[ {#1}]}
\renewcommand{\Im}[1]{{\rm Im}[ {#1}]}

\begin{document}


\title{Coulomb blockade for an oscillating tunnel junction }

\author{N.~Pauget}
\author{F.~Pistolesi}
\affiliation{
Universit\'e Joseph Fourier, Laboratoire de Physique et Mod\'elisation des Milieux Condens\'es,\\
C.N.R.S. B.P. 166, 38042 Grenoble, France
}

\author{M.~Houzet}
\affiliation{
INAC/SPSMS,
CEA Grenoble, 17, rue des Martyrs, 38054 Grenoble Cedex 9, France\\
}

\date{\today}

\begin{abstract}
We consider a tunnel junction formed between a fixed electrode and an oscillating one.
Accumulation of the charge on the junction capacitor induces a force on the nano-mechanical
oscillator.
The junction is voltage biased and connected in series with an impedance $Z(\omega)$.
We discuss how the picture of Coulomb blockade is modified by the presence of the oscillator.
Quantum fluctuations of the mechanical oscillator modify the $I$-$V$ characteristics
particularly in the strong Coulomb blockade limit.
We show that the oscillator can be taken into account by a simple modification of the effective
impedance of the circuit.
We discuss in some details the case of a single inductance $Z(\omega)=iL\omega$ and of a
constant resistance $Z(\omega)=R$.
With little modifications the theory applies also to incoherent transport
in Josephson junctions in the tunneling limit.
\end{abstract}


\maketitle


\section{Introduction}

The improved ability to build electronic devices on the nanometer scale opens the perspective
to exploit and study the coupling between electronic transport and mechanical degrees of
freedom.\cite{roukes:2001,blencowe:2004}
The elementary system typically considered in nano-electromechanics is the harmonic oscillator coupled
in some way to the electronic degrees of freedom.
A widely studied device is the single-electron transistor which consists in a small metallic island
connected to the leads through two tunnel barriers.
The oscillating part of the device can be either a nearby gate electrode, or the island itself.
The motion of the device modifies mainly two quantities: the capacitances and the (bare) tunneling rates.
According to which dependence dominates, several effects have been predicted or observed.
When only the capacitance is modified by the oscillation, it has been shown in the weak coupling limit
that the variance of the oscillator position satisfies the equipartition theorem
with the voltage bias replacing the temperature.\cite{armour:2004}
Under specific conditions, the system may also undergo dynamical instabilities.\cite{blanter:2004E,clerk:2005,usmani:2006}
In the strong coupling limit, a new kind of current blockade (Frank-Condon blockade) has been
predicted.\cite{braig:2003,mitra:2004,koch:2005,koch:2006}
This effect also persists in the classical limit.\cite{mozyrsky:2006,doiron:2006,pistolesi:2007}
When the oscillation modifies the distance between the metallic leads, it has been shown that
a dynamical instability can occur.\cite{gorelik:1998,gorelik:2001,armour:2002,novotny:2003,pistolesi:2005,jonsson:2005,pistolesi:2006}
This instability (called shuttle instability) is characterized by a synchronization of the electronic transport with the
mechanical oscillations, as is clearly seen from the full counting statistics of charge transport.\cite{pistolesiFCS:2004,romito:2004}
Observation of shuttling is difficult, since the distance between the metallic leads must be comparable with the
tunneling length in order to produce a measurable current.\cite{park:2000,erbe:2001,scheible:2004,pasupathy:2005}
Very recently indications of shuttling have been found by the authors of Ref. \onlinecite{kim:2007} probably in
an intermediate regime between tunneling and field emission.
The wide interest for the single electron transistor comes from its high sensitivity to a small variation of the gate voltage.
Thus, it can be used, for instance, as a sensitive displacement detector\cite{lahaye:2004} and it has been shown to
reach the quantum back action limit.\cite{naik:2006}

An even simpler device is the single tunnel junction where the distance between the two metals constitutes
the mechanical degree of freedom.
The effect of the position dependence of the tunneling resistance has been already considered in the
literature,\cite{schwabe:1995,clerk:2004,wabnig:2007} both for the current and the noise of the device.
In this case, the force acting on the oscillator originates from the electron-momentum transfer.
Recently, the position fluctuation has been detected by measuring the current fluctuations.\cite{flowers:2007}
This experiment probed the back action of the current crossing the tunnel junction on the oscillator.
Apparently, the position dependence of the tunneling matrix elements
is not sufficient to explain the intensity of the back action.
To our knowledge, the effect of the force coming from the variation of the capacitance $C$ as a function
of the distance between the leads has not been considered in this context so far.
A particularly interesting case is when the device is voltage biased in series with an impedance
$Z(\omega)$ leading, under certain conditions, to the Coulomb blockade physics.\cite{ingold:1992}
Then, similarly to what happens for the single electron transistor, one can expect that
a small position dependence of the capacitance can affect the current-voltage
characteristics of the device.

In this paper, we consider the effect of the mechanical oscillator on the $I$-$V$ curve.
We will work within the assumptions of the standard Coulomb blockade theory.\cite{ingold:1992}
Specifically, we assume that the tunneling resistance is sufficiently large so that the oscillator and the
electromagnetic environment have time to relax to equilibrium between two tunneling events.
This is the opposite limit with respect to the one considered in Refs. \onlinecite{clerk:2004,wabnig:2007} for the
tunneling dependent Hamiltonian.
There, the tunneling is so frequent that the system may reach a stationary off-equilibrium state.
Restricting to the linear coupling of the oscillator with the electromagnetic modes, we will discuss the effect
of the quantum fluctuations of the mechanical oscillator on the current.

The paper is organized as follows.
In Sect. \ref{sec2} we describe the model and its range of validity.
We show that within a linear approximation the mechanical oscillator
can be accounted by an effective impedance.
In Sect. \ref{sec3} we consider two specific cases of $Z(\omega)$ and
discuss the resulting conductance at zero temperature.
In Sect. \ref{sec4} we  consider the case of an oscillating
Josephson junction by extending the results of previous sections
to the superconducting case.
Section \ref{sec5} gives our conclusions.


\section{Model}
\label{sec2}

\begin{figure}
\includegraphics[width=0.9\columnwidth]{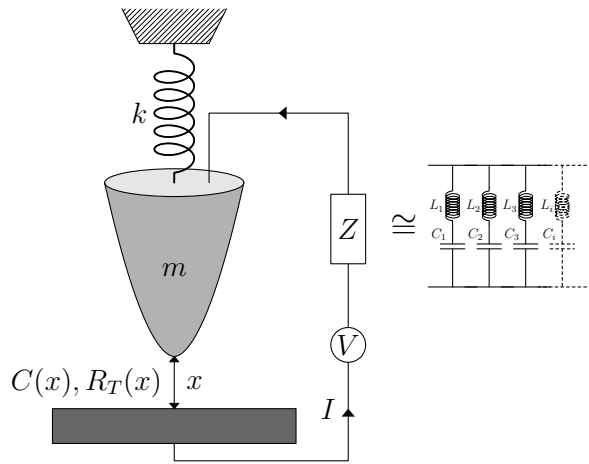}
\caption{
\label{Montagea}
Schematic representation of the system.
A tunneling contact is formed between an oscillating electrode
(here depicted as a STM tip) and a fixed surface.
The oscillator is modelled by
a spring constant $k$ and an effective mass $m$.
The junction between the tip and the plate is characterized by an effective
capacitance $C(x)$ and resistance $R_T(x)$, where $x$ denotes the position of the tip.
The circuit is closed by an impedance $Z$ and a voltage source $V$.}
\end{figure}

We consider the electronic transport through a circuit composed of
a tunnel junction in series with an arbitrary impedance $Z(\omega)$ and biased with
an ideal voltage source $V$.
We assume that one of the two metallic leads forming the tunnel junction is free to oscillate
and can thus modify its distance with respect the other one.
The single degree of freedom describing the oscillation is given by
$x$, the oscillator position.
Its effective mass and spring constant are $m$ and $k$, respectively.
The tunneling surface is very small.
Thus, it is important to account for the effective capacitor forming at the tunnel barrier.
Its capacitance $C(x)$ depends on the position of the oscillator.
The tunneling resistance $R_T(x)$ may also depend on $x$.
In the following we will neglect this dependence by showing that its effect is negligible
with respect to that of $C(x)$ in the regime of rare tunneling events that we are considering.
The device is depicted schematically in Fig. \ref{Montagea}.

%
%
%

\subsection{Time scales}

As anticipated in the Introduction we will work within the same hypothesis
that are commonly  used to describe Coulomb blockade.
Specifically, we assume that the relaxation times for the electromagnetic environment,
$1/\gamma_{RC}=RC$, and for the oscillator, $1/\gamma$, are shorter than the
average time between two tunneling events, $\tau_T \sim eR_T/V$.
Here, $R \sim \Re{Z}$ is the typical resistance associated with
the impedance $Z$ and $\gamma$ is the damping coefficient of the mechanical oscillator with frequency
$\omega_m=\sqrt{k/m}$.
If $1/\gamma_{RC},1/\gamma \ll \tau_T$, the system has time to come back to the equilibrium state
after an electron has crossed the tunnel barrier.
This allows us to calculate the average current on the equilibrium
quantum state of the environment.
It gives the following limitations on the voltage: $eV \ll E_C (R_T/R)$
and $eV\ll e^2\gamma R_T$, where $E_C=e^2/2C$ is the Coulomb energy.
Since we are interested in the range of voltages around $E_C/e$ we assume that
$R_T \gg R$ and $R_T \gg 1/C\gamma$.
The first condition is the standard tunneling condition, the second involves the
mechanical damping.
One should remember that the damping of the oscillator also
depends on the impedance of the circuit.
Thus, even if the oscillator is not
damped by a mechanical source, its quality factor would not be infinite.
In the limit of large $R_T$ the above conditions can thus always be satisfied.

Let us now discuss the effect of the position dependence of the
tunneling resistance.
To evaluate the importance of this effect for the dynamics of the
junction we consider the effective friction coefficient generated
by the position dependence of $R_T$:\cite{clerk:2004}
\beq
    \gamma_T \simeq \frac{R_Q}{R_T} \left( \frac{x_o}{\lambda_T} \right)^2 \omega_m
\eeq
where $R_Q=2\pi \hbar/e^2$ is the quantum of resistance, $x_o^2=\hbar/2m\omega_m$
is the (square of the) zero point motion amplitude of the oscillator and
$\lambda_T$ is the tunneling length defined by $\lambda_T^{-1}=(dR_T/dx)/R_T$.
The explicit expression of the friction coefficient induced by the electromagnetic
dissipation is given below for a specific case [cf. \refE{gammap}].
Independently on its explicit form, what matters is that
it does not depend on the tunneling
resistance since the damping is due to the environment and the capacitive
coupling.
Thus the ratio $\gamma_T/\gamma$ can always be made small by choosing
a large enough tunneling resistance.
Then we may safely neglect the position dependence of the tunneling matrix elements.
We will come back to this point at the end of Section \ref{sec3}.


\subsection{Derivation of the general expression for the current}

The Hamiltonian describing the system is the following:
\beq
    H = H_j+H_{osc}+H_{EM}.
\label{Hamiltonianfull}
\eeq
where
$
    H_j =
\sum_{k} \xi_{k} c^{\dagger}_{k} c_{k}
+\sum_{q} \xi_{q} c^{\dagger}_{q} c_{q}
+ H_T
$
with
$H_T = A+A^\dag$ and  $ A = \sum_{k,q} t_{kq} e^{i \phi(t) } c^{\dagger}_{k} c_{q}$.
Here, $k$ and $q$ label the femionic eigenstates in the fixed and oscillating
electrode, respectively, $c_k$ is a destruction operator in state $k$, with energy $\xi_k$,
and $t_{k q}$ the tunneling matrix element.
The phase $\phi(t)$ is obtained by integrating the voltage difference at the junction:
$
    \phi(t)= \int_{-\infty}^t { (eV_J(t')/\hbar})  \dd t'
$.
For a constant voltage biased junction $\phi(t)=eVt/\hbar$.
However, due to the impedance in series with the junction, the
voltage fluctuates and, in general, $V_J$ is different from $V$.
The Coulomb blockade physics stems from the behavior of the phase $\phi$ and,
specifically, from its correlation function.
The current through the junction can be obtained by perturbation theory in
$H_T$.
The operator for the current from the mobile electrode to the
fixed one reads
$
    I =- ie(A-A^\dag)/\hbar.
$
Linear response theory gives:
\beqa
    && I = {e\over \hbar^2} \sum_{k q} |t_{kq}|^2 \int_{-\infty}^{+\infty} \dd t e^{-i(\xi_k-\xi_q)t/\hbar}
        \times \nonumber \\
        &&
    \left[\qav{e^{-i\phi(t)} e^{i \phi(0)}} f_k(1-f_q)-
    \qav{e^{i\phi(0)} e^{-i \phi(t)}} f_q(1-f_k)\right]
    \nonumber\\
    &&
    \label{currentPhi}
\eeqa
where $f_{k,q}=1/(e^{\beta \xi_{k,q}}+1)$ are the Fermi distributions in the leads,
and $\beta$ is the inverse temperature.
In order to include the quantum fluctuation of the electromagnetic fields, the
phases in this expression must be regarded as operators whose dynamics is determined
by the rest of the Hamiltonian \refe{Hamiltonianfull}.
We will restrict to the quadratic part of the Hamiltonian.
Within this approximation,
the fields $\phi$ satisfy Wick theorem and expression \refe{currentPhi} can be simplified
by making use of the exact relation:
\beq
    \qav{e^{i\tilde \phi(t)} e^{-i  \tilde \phi(0)} }
    =
   e^{J(t)-J(0)}
\eeq
where $\tilde \phi=\phi(t)-eV t/\hbar$ and
$
    J(t)= \qav{\tilde \phi(t)\tilde \phi(0)} 
$.

The current can then be expressed in terms of the Fourier transform of
$e^{ J(t)-J(0)}$:
\beq
    P(E) = \int_{-\infty}^{+\infty} {\dd t \over 2 \pi \hbar} e^{J(t)-J(0)+ i E t/\hbar}
    \,.
    \label{PofE}
\eeq
The function $P(E)$ gives the probability that the environment absorbs a quantity of energy equal to $E$ during the tunneling process.
Finally the current in terms of $P(E)$ reads:\cite{ingold:1992}
\beq
    I = {1 \over e R_T}  \int_{-\infty}^{+\infty} E \dd E P(eV-E) { (1-e^{-\beta e V})\over 1 - e^{-\beta E}}
    \,,
    \label{current}
\eeq
where $R_T$ is the tunneling resistance in the absence of the environment [{\em i.e.} when $P(E)=\delta(E)$].


In order to find the behavior of the device we need to determine the function $P(E)$
in the presence of the oscillator.
Since we will restrict to the quadratic part of the Hamiltonian, we can obtain the
quantum phase correlation function $J(t)$ from the classical response function
of the phase to a current source at the junction.
Specifically we add to the total Hamiltonian \refe{Hamiltonianfull} the term
$
    -I_{ext}(t) \tilde{\phi}_{\mathrm{int}} (t)\hbar/e
$,
where $\tilde{\phi}_{\mathrm{int}} (t)$ is the phase in the interacting picture.
The linear response to the external field $I_{ext}$ gives
\beq
    \qav{\tilde \phi(t)} = -\int_{-\infty}^{+\infty} \dd t' \chi(t-t') I_{ext}(t') {\hbar \over e}
\eeq
with
$
    \chi(t)={i\over \hbar} \qav{[\tilde{\phi}_{\mathrm{int}} (0),\tilde{\phi}_{\mathrm{int}} (t)]   } \theta(t)
$.
Using the fluctuation-dissipation theorem one can show that
\beq
    J(\omega)=-2 \hbar {\Im{\chi(\omega)} \over  1-e^{\beta \omega}}
    \,.
    \label{Cofomega}
\eeq
Since the Hamiltonian is quadratic, $\chi(\omega)$ can be
calculated from the classical response function (see for instance Ref. \onlinecite{weiss:2000}).
The problem is now reduced to the determination of the classical retarded response
function.
This can be readily obtained by solving the classical equations of motion
by Laplace transform.


We determine now the equations of motion for the phase, the position of the oscillator,
and the degrees of freedom of the environment.
To describe the dissipation we introduce a bath of harmonic oscillators coupled to the
mechanical oscillator and a bath of electromagnetic oscillators coupled to the
current through the tunnel junction.
The coupling of the mechanical and electrical part comes from the charge energy
$Q^2/2C(x)$ of the capacitor.
The equations of motion read:
\beqa
    m \ddot x &=& -kx   -\sum_p k_p (x-x_p) + {1 \over 2} C'(x) \dot \phi ^2\left({\hbar \over e}\right)^2
    \label{S1}
    \\
     C(x) \ddot \phi &=& -C'(x) \dot x \dot \phi +\sum_n  {1 \over L_n}(\phi_e-\phi_n-\phi)
     + I_{ext}(t){e\over \hbar}
     \nonumber \\
     &&
    \label{S2}
    \\
    m_p \ddot x_p &=& k_p (x-x_p)  \label{S3}
    \\
    C_n \ddot \phi_n  &=& {1 \over L_n}(\phi_e-\phi_n-\phi)  \label{S4}
    \,.
\eeqa
Here $x_p$ and $\phi_n$ are the degrees of freedom of the environment,
$\phi_e(t)=eV t/\hbar$ is due to the constant external voltage source and we choose the zero of
$x$ as the equilibrium position of the oscillator when $V=0$.
We also define $C'(x)=dC/dx$.
For $I_{ext}=0$, this system of equations has a stationary solution given by $x=x_{eq}=C'(x_{eq})V^2/2k$,
$\dot\phi=eV/\hbar$, $x_p=x_{eq}$ and $\phi_n=0$.
This solution is unique for a given gauge.
We thus expand the equations around this solution and consider the quadratic fluctuations of the fields only:
$x(t)=x_{eq}+\tilde x(t)$, $x_p(t)=x_{eq}+\tilde x_p(t)$, and $\phi(t)=eVt/\hbar+\tilde \phi(t)$.
Let us introduce the Laplace transform for the fluctuating fields.
For example, for $x$ it reads:
$\tilde x(s) = \int_0^{+\infty} dt e^{-s t} \tilde x(t)$, with $\Re s>0$
and
$\tilde x(t)=\int_{-i\infty+a}^{+i\infty+a} (ds/2\pi i) e^{s t} \tilde x(s)$
with $a>0$.
The system of differential equations then becomes an algebraic system of equations.
We are interested in the response function.
Thus, we can neglect transient terms.
Keeping only linear terms in the fluctuating fields, and solving first for the
degrees of freedom of the environment one finds:
\beqa
    \left[s^2 + \gamma(s)s+\omega_m^2\right] \tilde x &=&   {C' V \hbar \over e m} s \tilde \phi
    \label{SSS1}
    \\
    \left[C s + Y(s)\right]s \tilde\phi &=& -C' s \tilde x {eV\over \hbar} + I_{ext}(t){e\over \hbar}
    \label{SSS2}
    \,,
\eeqa
where we have introduced the memory function of the oscillator and the impedance of the circuit:
\[
    \gamma(s)= {1\over m} \sum_p k_p {s \over s^2+\omega_p^2},
    \quad
    Z(s/i)^{-1} =  \sum_n {1 \over L_n} {s \over s^2+\omega_n^2}.
\]
The two coupled equations \refe{SSS1} and \refe{SSS2} describe the response of the phase
$\tilde \phi$ to the external current $I_{ext}$ flowing through the tunnel junction.
Solving for $\tilde \phi$ and letting $s\rightarrow i \omega+0^+$ we obtain the expression for
$\chi(\omega)$.
Using \refE{Cofomega}, the expression for $J(\omega)$ reads:
\beq
    J(\omega) = {4\pi\over\omega}  {\Re{Z_t(\omega)}\over R_Q} {1\over 1-e^{\beta \omega}}
    \label{usufulCom}
\eeq
with
\beq
    Z_t(\omega) =
    \left[i\omega  C + Z(\omega)^{-1} +
        {{C'}^2 V^2 i \omega /m \over  \omega_m^2 + \gamma(i\omega) i\omega -\omega^2}\right]^{-1}
        \label{response}
        \,.
\eeq
The usual result of the Coulomb blockade can be obtained by setting $C'=0$ in this
expression.
Then, only the electromagnetic environment would contribute to the response function.
The presence of the oscillator simply modifies the response of the system; for $\gamma$ constant, it would be
equivalent to a RLC circuit in parallel with the tunnel junction.

The current voltage characteristics can now be obtained by combining Eqs. \refe{PofE},
 \refe{current}, \refe{usufulCom}, and  \refe{response}.
In the following section we consider two representative cases for the circuit impedance
and discuss the expected conductances.

\section{Results in specific cases}
\label{sec3}

We consider now two simple cases for the impedance of the circuit:
$Z=i\omega L$ and $Z= R$.
In the first case (Sect. \ref{subsectionInductive}), the electromagnetic environment is an LC-oscillator.
Thus, the system reduces to two coupled oscillators: one mechanical and the other electromagnetic.
This example is interesting for his simplicity.
It clearly shows how the mechanical
and electric part becomes coupled and how this coupling appears in the
current voltage characteristics.
By contrast, its experimental realization is non-trivial, since it is in general
not easy to get rid of parasitic capacitances and resistances.
An inductance and a capacitance are present in the environment for the oscillating
tunneling junction in the experiment of Ref. \onlinecite{flowers:2007}.
The environment forms in this way a radio frequency resonator that is used
to improve the band-width of the detection scheme.
However, one should note that experiment of Ref. \onlinecite{flowers:2007}
is not in the regime of parameters considered in the present paper.

In the second case (Sect. \ref{subsectionohmic}), the environment provides the
dissipation and the conditions
for the standard Coulomb blockade with the suppression of the current at low
voltages.
This case is more relevant from the experimental point of view, since high impedance
environment can be more easily devised.
For electronic transport without mechanical motion  an accurate experimental study of the effect of a
purely resistive environment has been carried out recently by tuning the environment resistance for a given
tunnel junction.\cite{altimiras:2007}

\subsection{Inductive case $Z(\omega)=iL\omega$}
\label{subsectionInductive}


We consider the case of undamped harmonic oscillator.
Strictly speaking, the hypothesis of thermal equilibrium for the
oscillator and for the electromagnetic modes is not valid.
We will assume that an infinitesimal dissipation is present and that the
tunneling resistance is sufficiently large to let the time for the system to
relax between two tunneling events.
The electromagnetic part of the circuit is characterized by the oscillator
frequency $\omega_{LC}=1/\sqrt{LC}$.
The pure electromagnetic case is discussed in the review paper Ref.~\onlinecite{ingold:1992}.
Due to the coupling with the mechanical mode two resonant frequencies appear,
as can be seen by finding the poles of $Z_t(\omega)$ [cfr. \refE{response}]:
\beq
    \omega_\pm^2 = \frac{\left[\omega_m^2+\omega_{LC}^2+\omega_I^2 \pm
        \sqrt{(\omega_{LC}^2+\omega_m^2+\omega_I^2)^2-4\omega_{LC}^2 \omega_m^2}\right]}{2}.
\eeq
Here we introduced the ``coupling frequency'' $\omega_I^2=C'^2 V^2 /Cm$.
The real part of $Z_t$ reduces to a sum of four delta functions
\beq
    {\Re{Z_t(\omega)}\over R_Q} = \sum_{\sigma=\pm} \frac{\rho_\sigma \omega_\sigma}{2}
    \left[\delta(\omega-\omega_\sigma)+\delta(\omega+\omega_\sigma)\right]
\eeq
with
$
    \rho_{\pm}
    =
    ({E_C}/{\hbar \omega_{\pm}}) | {\omega_{\pm}^2 - \omega_m^2}|/|{\omega_{\pm}^2 - \omega_{\mp}^2}|
$.

It is convenient to express the coupling of mechanical and electronic
degrees of freedom in terms of a parameter that depends only on the device
and does not depend on the voltage bias.
The reason is that it is easy to vary experimentally
the voltage for a given device.
A physically relevant parameter is the variation of the elastic energy $E_e$
when the oscillator is displaced by a distance $\Delta x=E_C(C'/C)/k$ in response
to the Coulomb force generated by a single electron on the capacitor.
With this definition we have
\beq
    E_e=E_C^2(C'/C)^2/2k
    \quad
    {\rm and} \quad
    \omega_I^2 = \omega_m^2 E_e (eV)^2/E_C^3
    \,.
\eeq

For a junction made of a STM tip on top of a suspended carbon nanotube, an electromechanical coupling with
the radial breathing mode of the tube was reported \cite{leroy:2004-2,leroy:2005}.
In this context, $\omega_m$ and $E_c$ were of the same order of magnitude (around $10meV$).
Thus, we estimate the coupling parameter $E_e/E_c\sim (x_0/L_c)^2$, where $x_0$ is the zero-point motion
 of the mode (a fraction of Angstr\"om) and $L_c=C/C'$ (a fraction of nm).
In the following, we assume a (rather optimistic) ratio of $E_e/E_c\lesssim 0.1$

For small value of $E_e/E_C$ and $\omega_m\neq \omega_{LC}$ the new frequencies $\omega_m'$ and $\omega_{LC}'$
are only weakly modified by the interaction with respect to their bare values:
\beqa
    \omega_m'^2/\omega_m^2 &=&1+{\omega_I^2/(\omega_m^2-\omega_{LC}^2})
    \\
    \omega_{LC}'^2/\omega_{LC}^2 &=& 1-{\omega_I^2/(\omega_m^2-\omega_{LC}^2}) \label{Ee}
\eeqa
(note that $\omega_m'=\omega_+$ if $\omega_m>\omega_{LC}$ and $\omega_m'=\omega_-$ if
$\omega_m<\omega_{LC}$)
and the weights of the poles simplify:
\beq
    \rho_m \approx  {E_C\over \hbar \omega_m} {\omega_I^2\omega_m^2 \over (\omega_{LC}^2-\omega_m^2)^2} \ll
    \rho_{LC} \approx {E_C\over \hbar \omega_{LC}}.
\eeq
The inequality holds if the bare frequencies $\omega_m$ and $\omega_{LC}$ are of the same
order of magnitude.
On the other hand, when $\omega_m=\omega_{LC}\gg \omega_I$, we find $\omega_\pm=\omega_m \pm \omega_I/2$
and $\rho_+=\rho_-=\rho_{LC}/2$.
\begin{figure}
\begin{center}
\includegraphics*[width=\columnwidth]{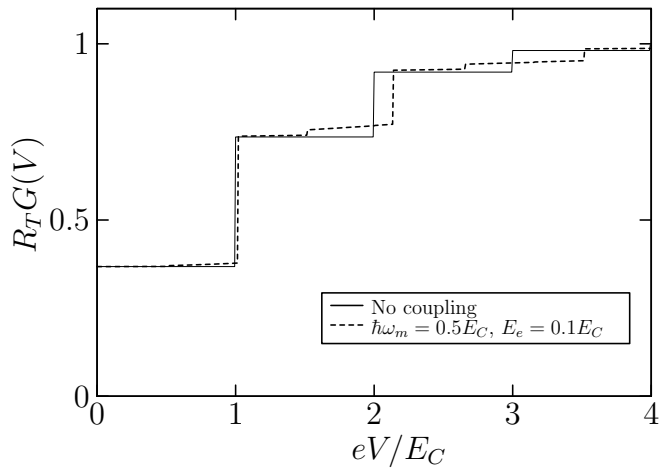}
\caption{
Differential conductance of the device as a function of the bias voltage for the case $Z=i\omega L$.
The bare oscillating frequencies are $\hbar \omega_m=0.5 E_C$ and $\hbar \omega_{LC}=E_C$.
The electromechanical coupling is $E_e=0$ (full line) and $E_e = 0.1 E_C$ (dashed line) (in this case,
signatures of mechanical coupling are not visible for $\hbar \omega_m$ below $0.5E_C$).
}
\label{fig4}
\end{center}
\end{figure}

>From \refE{usufulCom}, one can obtain $J(t)-J(0)=\sum_\sigma \rho_\sigma j_\sigma(t)$
with:\cite{ingold:1992}
\beq
    j_{\sigma} (t)
    =
    \coth \frac{\beta \hbar \omega_{\sigma}}{2} \left( \cos (\omega_{\sigma} t) - 1\right)
    -i \sin (\omega_{\sigma} t)
    \,.
\eeq
We consider now the zero temperature limit.
In this case, the function $P(E)$ obtained from
 \refE{PofE} has a simple expression in terms of an infinite series:
\beq
    P(E) = \sum_{n,n'=0}^{\infty}
    p_{n}(\rho_m) p_{n'}(\rho_{LC})
    \delta (E- n \hbar \omega'_m - n'\hbar \omega'_{LC})
\eeq
where $p_n(\rho)=e^{-\rho} \rho^n/n!$.

The differential conductance $G=dI/dV$ can be obtained from the current \refe{current}.
Neglecting the weak $V$-dependence of $\rho_\sigma$ and $\omega_\sigma$ in the derivative
we obtain:
\beq
    R_T G(V) = \sum_{\hbar n \omega'_m  + \hbar n'\omega'_{LC} < eV} p_{n}(\rho_m) p_{n'}(\rho_{LC})
    \,.
\eeq
It leads to a double series of steps spaced by $\hbar \omega_m$ and $\hbar \omega_{LC}$.
If $\rho_m=0$, one recovers the usual sequence of steps with a suppression of the
conductance at low voltage and $R_T G=1$ for $eV \gg E_C$.
The interesting situation is when $\rho_{LC}$ is of the order of $1$.
In this case steps are
clearly visible for $eV \sim E_C$ (cf Fig. \ref{fig4}).
If $\rho_m \neq 0$ and $\omega_m<\omega_{LC}$ the mechanical oscillator
manifests itself as a series of additional steps of width $\hbar \omega'_m$ particularly visible
at the beginning of each step of the LC circuit.
In practice the picture is slightly more complex due to the dependence on $V$ of the four
parameters $\rho_\sigma$ and $\omega_\sigma$.
The renormalization of the frequency and the change of $\rho$ introduces a shift of the
steps and a weak smooth $V$-dependence of the conductance plateaux, as shown in Fig. \ref{fig4}.
The presence of the mechanical oscillator thus gives a similar result to what has been
predicted for a single electron transistor\cite{braig:2003} with Frank-Condon steps
and a reduction of the current at low voltage.
This result is found here in a much simpler structure: a single tunnel
junction coupled in series with an external inductance.


\subsection{Ohmic case $Z(\omega)=R$}
\label{subsectionohmic}

Let us now consider the pure Ohmic case for the external impedance
$Z(\omega)=R$.
The first step is again to find the poles of $Z_t(\omega)$.
For $\gamma(\omega)$ constant,
three poles are present: one is pure imaginary, $i \gamma_{RC}'$, and the two other ones are
complex, $\omega_{\pm} = \pm \omega_m' +i \gamma'/2$.
In the relevant limit of small coupling $\omega_I \ll \omega_m, \gamma_{RC}$ and low mechanical damping $\gamma\ll \omega_m, \gamma_{RC}$ we get:
\beqa
    \gamma_{RC}' &=& \gamma_{RC}- {\omega_I^2\gamma_{RC} \over \gamma_{RC}^2+\omega_m^2 }+\dots\,,
    \\
    \omega_m'    &=& \omega_m+\frac12 {\omega_m \omega_I^2 \over \gamma_{RC}^2+\omega_m^2} - \frac{\gamma^2}{8\omega_m}+\dots ,
    \\
    \gamma'     &=& \gamma +  {\gamma_{RC} \omega_I^2 \over \gamma_{RC}^2+\omega_m^2}+ \dots
    \label{gammap}
    \quad .
\eeqa
The mechanical oscillator acquires a damping even if $\gamma=0$.
This intrinsic damping is proportional to the coupling.
>From here, we drop the bare mechanical damping since it does not induce new effects compared to those
already introduced by the presence of $R$.
An important parameter for this device is the dimensionless conductance of the tunnel
junction $g=R_Q/R$.
For $g \gg 1$ the Coulomb blockade is no more observable, since the charge can vary
continuously.
In the absence of the oscillator the appearance of the Coulomb
blockade is signaled by the change of the low-energy behavior of
$P(E)$, that goes like $E^{2/g-1}$.\cite{ingold:1992}
Thus for $g < 2$ the current is suppressed at low voltage $V < E_C/e$.
If we consider the interesting case of $\hbar \omega_m \sim E_C$,
then the condition $g \ll 1$ implies also that $\gamma_{RC} \ll \omega_m$.
In this limit the real part of $Z_t$ is simply the sum of three
Lorentzian functions:
\beqa
    &&{\Re{Z_t(\omega)} \over R_Q}
    =  {E_C\over \hbar}
    \left[ \epsilon L(\omega,2 \gamma_{RC}') \right.
    \nonumber \\
    &&+  \left.
     {\gamma' \over \gamma_{RC}} \left[ L(\omega-\omega_m',\gamma') +L(\omega+\omega_m',\gamma')\right]
    \right] \label{realZ}
\eeqa
with $\epsilon={\gamma_{RC}'/\gamma_{RC} }$, and $L(\omega,\gamma)=(\gamma/2)/(\omega^2+\gamma^2/4)/\pi$.
Since the coupling is weak, the damping of the mechanical mode is
also very small and the Lorentzian function can be approximated by
a Dirac delta-function [$L(\omega,\gamma\rightarrow 0)=\delta(\omega)$].
At zero temperature this gives for the phase correlation function:
\beq
    J(t)-J(0)=  {E_C \over \pi \hbar \gamma_{RC} } j_o(\gamma_{RC}' t)+ \rho_m[e^{-i\omega_m't}-1]
\eeq
where $\rho_m=\frac{2E_C}{\hbar \omega_m} \frac{\gamma'}{\gamma_{RC}}$ and
$j_o(\tau)= 2\int_0^{\infty} {e^{i x \tau}-1 \over x(x^2+1)} \dd x$
is the correlation function of the circuit in the absence of the
mechanical oscillator.
$P(E)$ thus reads:
\beq
    P(E)=\sum_{n=0}^\infty {p_n(\rho_m) \over \epsilon} P_o[(E-n\hbar \omega_m')/\epsilon]
    \label{PofEfinal}
\eeq
where
\beq
    P_o(E) = \frac{1}{2 \pi \hbar} \int_{-\infty}^{+\infty} \dd t e^{i E t/\hbar + (E_C /\pi \hbar \gamma_{RC})j_o(t\gamma_{RC})}
\eeq
is the function $P(E)$ in the absence of the oscillator.
For the case at hand of small $g$ the function $P_o(E)$ is
peaked at $E\approx E_C$ with a width $\sim (E_C/\pi) \sqrt{2 g|\log g|}$
that vanishes for $g\rightarrow 0$.
Thus, the resulting $P(E)$ is a sequence of peaks shifted by
$\hbar \omega_m$.
\begin{figure}
\begin{center}
\includegraphics*[width=\columnwidth]{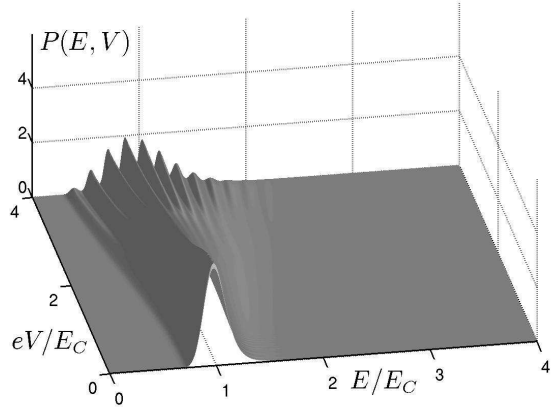}
\caption{$P(E)$ is displayed for different values of $V$.
We take $g=0.015$, $\hbar \omega_m=0.1 E_C$ and $E_e=0.1 E_C$.
}
\label{figRes1}
\end{center}
\end{figure}
Figure \ref{figRes1} shows the dependence of $P(E)$ on $E$ and
$V$ as obtained by numerical integration.
For $eV/E_C \ll \sqrt{E_C/E_e}$ (Eqs. (\ref{Ee}) and (\ref{realZ})) the mechanical coupling is
negligible and one recovers the usual picture of Coulomb
blockade, with a single peak at $E\approx E_C$.
For larger values of $eV/E_C$ new peaks appear and
their positions slowly drift as a function of
$V$ towards higher energies.
The first peak instead drifts towards smaller energies, due to the
$\epsilon$ factor in \refe{PofEfinal}.
Since $\gamma_{RC}'$ reduces for larger values of $V$,
the width of the peaks also reduces.

\begin{figure}
\begin{center}
\includegraphics*[width=\columnwidth]{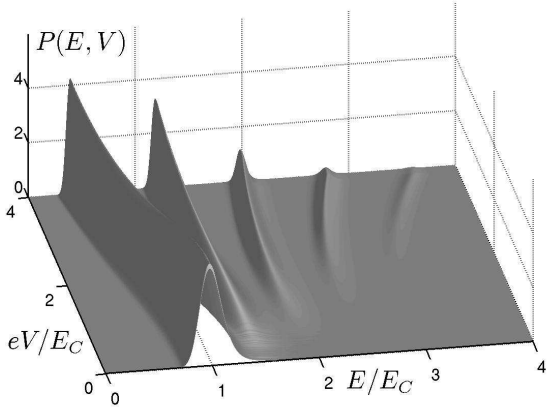}
\caption{
$P(E)$ is obtained numerically for different values of $V$ for
 $g=0.015$, $\hbar \omega_m=0.5 E_C$ and $E_e=0.1 E_C$.
}
\label{figRes2}
\end{center}
\end{figure}

In Fig. \ref{figRes2} we show the same function, but for a higher mechanical frequency $\hbar \omega_m/E_C=0.5$.
It shows the appearance of the peaks and their shrinking.

\begin{figure}
\begin{center}
\includegraphics*[width=\columnwidth]{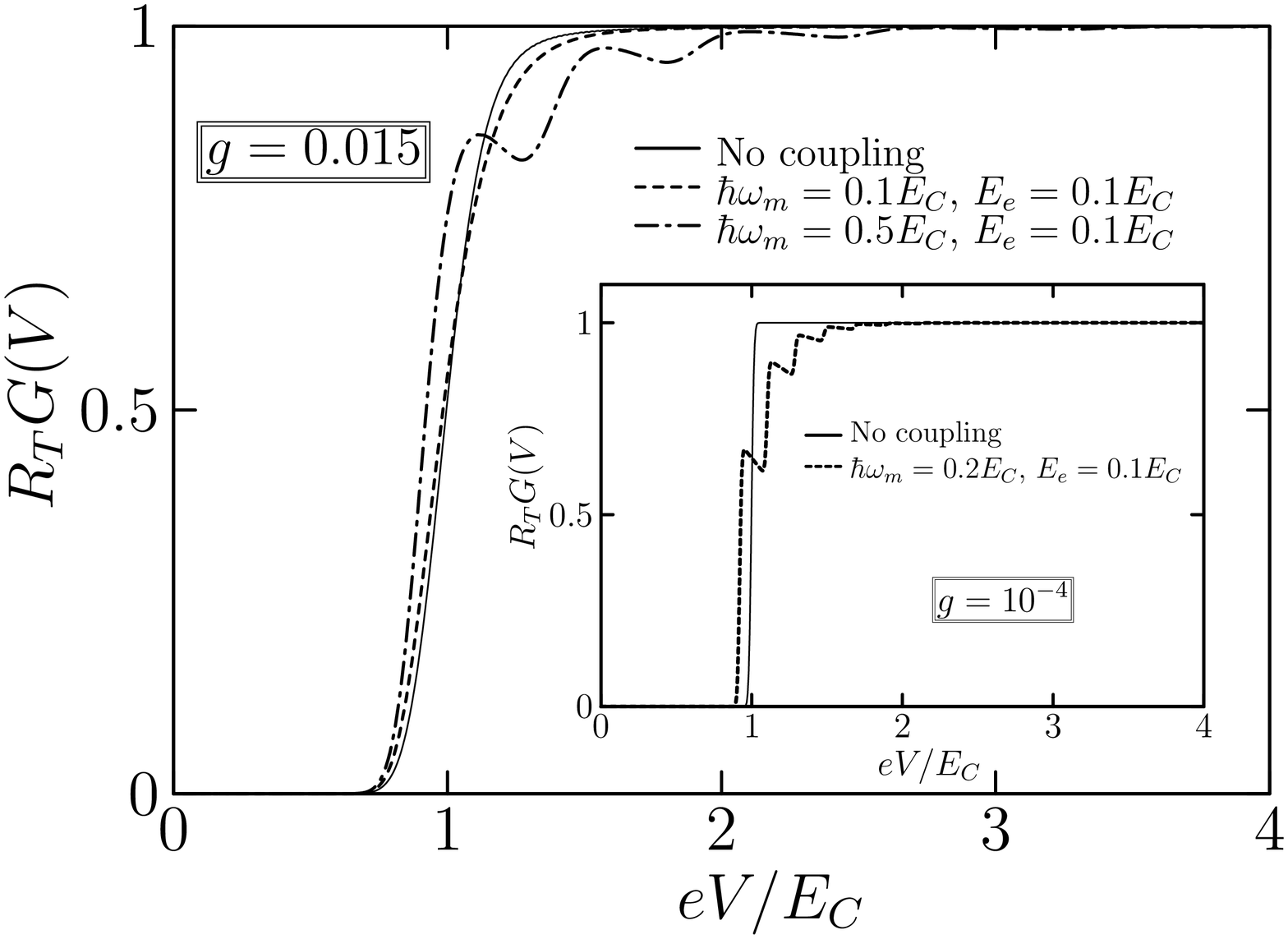}
\caption{
Main plot: $G(V)$ for the same parameters as in Fig.~(\ref{figRes1}) (dashed line) and Fig.~(\ref{figRes2})
(dot-dashed line) and for vanishing coupling (full line).
Inset: $G(V)$ for $g =10^{-4}$  from the analytical expression \refe{analyticalG} for $\hbar \omega_m = 0.1 E_C$,
$E_e = 0$ (full line) and $E_e = 0.1 E_C$ (dashed line).
The conductance is much more sensitive to the electromechanical coupling
in presence of a high impedance environment.
\label{figRes3}
}
\end{center}
\end{figure}

An important case is $g\rightarrow 0$, for which
$P_o(E)=\delta(E-E_C)$.
This is clearly the most favorable case to observe the quantum
fluctuations of the oscillator directly in the $I$-$V$ characteristics.
Indeed, even in the weak electromechanical coupling the sharpness of the
peaks gives rise to steps in $G(V)$, like in the case of a purely electromagnetic
resonator.
The conductance has thus the form (again neglecting the small
voltage dependence of $\rho_m$ and $\hbar \omega_m$ in performing
the derivative):
\beq
    R_T G(V)= \sum_{\hbar n \omega_m'<eV-\epsilon E_C }p_n(\rho_m)
    \label{analyticalG}
\eeq
the result is plotted in the insert of Fig. \ref{figRes3} for $E_e/E_C=0.1$ and $g=10^{-4}$.

For the general case the conductance can be obtained numerically
by integrating $P(E)$.
It is shown for three different values of the ratio $E_e/E_C$ in Fig. \ref{figRes3}.
The presence of the mechanical coupling introduces oscillations of the conductance
on the scale $eV \sim \hbar \omega_m$
that are precursors of the step behavior in the extreme $g=0$ case.
The small increase of the conductance at low voltage is
the consequence of the renormalisation of $E_C$ due to the
electromechanical coupling (the $\epsilon$ factor).

A last comment is in order for the resistive case.
Even if this system is described by a very similar electric circuit
to that used for a single electron transistor without gate, some differences
should not be overlooked.
In the single electron transistor the two resistance are tunnel junctions,
while here one is a tunnel junction and one a ohmic resistor.
A tunnel junction is substantially different from a ohmic resistor, since
in the first case the charge can be transferred only by a sudden tunneling event,
while in the second the charge slowly leaks through the resistor.
The charge quantization is thus not enforced by the Ohmic resistor,
and this has consequences also on the time dependence of the force
acting on the mobile part.
For these reasons our results complement those obtained by others authors
for suspended single electron transistors.\cite{armour:2004,blanter:2004E,clerk:2005,braig:2003,mitra:2004,koch:2005,mozyrsky:2006,doiron:2006,pistolesi:2007}

\subsection{Range of validity}\label{Rangevalidity}

Let us return to the comparison of position dependence of the tunneling and capacitive terms.
Using the explicit expression \refe{gammap} for $\gamma'$, we find that the condition
to neglect the effect of the tunneling force
(case $\gamma_{RC}\ll \omega_m$) is:
\beq
    1 \gg {\gamma_T\over \gamma'}
    =
    \left( {x_o \over \lambda_T} \right)^2  { \omega_m R_Q E_C^3\over \gamma_{RC} R_T E_e e^2 V^2}
    .
\eeq
This gives the explicit condition on the voltage in order to neglect the
position dependent part of the Hamiltonian.
Formally, it is clear that it can be always fulfilled for $R_T$ sufficiently large.

A second possible source of deviation from the behavior we found
comes from the cubic term of of the interaction energy.
We neglected the term $(\dot{\tilde \phi})^2$ with respect
to $eV \dot {\tilde \phi}/\hbar$.
This is justified when
$\langle (\dot { \tilde\phi})^2\rangle^{1/2} \ll eV/\hbar$.
We thus obtain the condition of validity of the quadratic theory at
zero temperature:
\beq
    {eV\over E_C} \gg \sqrt{g|\ln g|}
    \,.
\eeq

\section{Josephson junction}
\label{sec4}

Observing a Josephson supercurrent through
a STM tunnel junction has been shown to be feasible.\cite{naaman:2001}
Particularly appealing is the possibility of studying an oscillating
Josephson junction.
The effect of the position dependence of the electronic tunneling amplitude
on the dynamics of the Josephson junction has been considered recently.\cite{zhu:2006}
However the effect of the position dependence of the capacitance (Coulomb interaction)
has not been explored so far.
The theory of Coulomb blockade for normal metallic electrodes presented above can be generalized with little change to the superconducting case.
The main difference is that the phase difference is now a truly observable quantity and it is
conjugated with the number of Cooper pairs that cross the junction.
%

The Josephson junction is characterized by a critical current $I_c$ that defines the
energy scale $E_J=\hbar I_c/2e$.
For $E_J/E_C\ll 1$ a tunneling approach can be used to obtain the dissipative
current of the device.
We remind the main steps of the procedure.\cite{ingold:1992}
The current operator is
$
    I_S = I_c \sin (2\phi)
$
while the Josephson part of the Hamiltonian reads
$
    H_J = E_J \cos(2\phi)
$
(we keep the same definition of the phase, the factor two intervenes in the usual Josephson relation).

\begin{figure}
\begin{center}
\includegraphics*[width=\columnwidth]{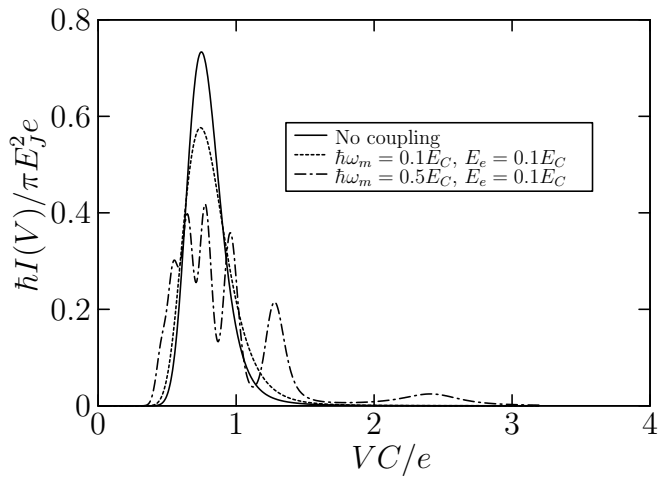}
\caption{$I(V)$ is plotted for $g=0.065$.
We distinguish the case $E_e = 0$ (full line), $E_e = 0.1 E_C$ with $\hbar \omega_m = 0.1 E_C$ (dashed line) and $E_e = 0.1 E_C$ with $\hbar \omega_m = 0.5 E_C$ (dot-dashed line).
As long as $\hbar \omega_m$ remains smaller than the width of the peak, the coupling manifests itself in a broadening of the Coulomb peak.
For $\hbar \omega_m/E_C = 0.5$, contributions of the mechanical oscillator clearly appear.
\label{figsupra}
}
\end{center}
\end{figure}

Calculating the current in the linear response limit we get
\beq
    I_S = I_c {E_J \over 4 \hbar}  \int_{-\infty}^{+\infty} \dd t' \qav{\left[e^{2 i \phi(t)}, e^{-2i \phi(t')}\right]} e^{\frac{2 i e V t}{\hbar}}
    \,.
\eeq
One can thus express the current again in terms of the Fourier transform of the phase
correlator:
\beq
    I(V) = {\pi E_J^2 e \over \hbar} \left[P_s(2eV)-P_s(-2eV) \right]
\eeq
where the definition of $P_s(E)$ differs from the definition of $P(E)$
simply by a factor of 4 in front of $J(t)$:
\beq
    P_s(E) = \int_{-\infty}^{+\infty} \dd t e^{4J(t)-4J(0)+i Et/\hbar}
    \,.
\eeq
The function $J(t)$ depends on the mechanical response function as
shown in Section \ref{sec2}.
The results of Sects. \ref{sec2} and \ref{sec3} can thus be applied to the superconducting
case to obtain $P_s(E)$ and hence $I(V)$.
In the Figure \ref{figsupra} we show the expected current-voltage characteristics
for an oscillating Josephson junction in the pure resistive case of Sect. \ref{sec3}.
The superconducting device is much more sensitive to the presence of the
mechanical oscillator, since the voltage dependence is directly related to
$P_s(E)$, and not to its integral.
One can then observe directly the Coulomb blockade peaks in the $I$-$V$ characteristics.

\section{Conclusion}
\label{sec5}

In this work, we considered the current-voltage characteristics of an oscillating tunnel junction.
At low tunnel coupling and moderately low voltage bias, we found that the coupling between the electronic and mechanic degrees of freedom mainly originates from the position dependence of the junction capacitance.
That is, the electrons crossing the tunnel junction couple to an electromagnetic environment formed by the rest of the circuit which is itself coupled to the mechanical oscillator.
In our study, we included both the electromagnetic and mechanical environments in the Coulomb blockade theory.
We found that the mechanical properties of the junction result in an additional mechanism for the current suppression that is similar to the Frank-Condon blockade predicted in an oscillating single-electron transistor.
As in the conventional Coulomb blockade, similar, though more pronounced, effects would also appear in a voltage-biased oscillating Josephson junction.

At larger tunneling rates, two new ingredients should be added to our study.
First, the electromagnetic and mechanical degrees of freedom would not have time to relax to equilibrium between two tunneling events.
Second, the position dependence of the tunneling matrix element may not give anymore negligible effects compared to those arising from the position dependence of the capacitance.
Considering both, it may be interesting to investigate whether a shuttling instability takes place in a single oscillating tunnel junction.

\section*{Acknowledgements}

We acknowledge useful discussions with D. Bagrets.
This work has been supported by the French {\em Agence Nationale
de la Recherche} under contract ANR-06-JCJC-036 NEMESIS.
F. P. thanks A. Buzdin and his group for hospitality at the { \em Centre de Physique
Moleculaire Optique et Hertzienne} of Bordeaux (France) where part of this work has
been completed.

\bibliography{redac,biblioNEMS}

\end{document}